# Food for Thought: Analyzing Public Opinion on the Supplemental Nutrition Assistance Program

Miriam Chappelka
University of Pennsylvania
Philadelphia, PA, USA
cmiriam@sas.upenn.edu

Jihwan Oh
Georgia Institute of Technology
Atlanta, GA, USA
Jihwan.Oh@gatech.edu

Dorris Scott
University of Georgia
Athens, GA, USA
dorris.scott25@uga.edu

Mizzani Walker-Holmes
Kennesaw State University
Marietta, GA, USA
mwalker@kennesaw.edu

## ABSTRACT
This project explores public opinion on the Supplemental Nutrition Assistance Program (SNAP) in news and social media outlets, and tracks elected representatives' voting records on issues relating to SNAP and food insecurity. We used machine learning, sentiment analysis, and text mining to analyze national and state level coverage of SNAP in order to gauge perceptions of the program over time across these outlets. Results indicate that the majority of news coverage has negative sentiment, more partisan news outlets have more extreme sentiment, and that clustering of negative reporting on SNAP occurs in the Midwest. Our final results and tools will be displayed in an on-line application that the ACFB Advocacy team can use to inform their communication to relevant stakeholders.

## 1.INTRODUCTION
The Supplemental Nutrition Assistance Program (SNAP), formerly known as food stamps, is a federal program that helps low income individuals purchase food. The Atlanta Community Food Bank (ACFB) aspires to eliminate hunger in its service area by 2025. To help achieve this goal, the food bank is raising awareness about the importance of SNAP. Their audience is stakeholders who contribute to the Atlanta Community Food Bank (who may be skeptical of the food bank's support of SNAP) and politicians (who can influence SNAP policy). We are assisting the food bank by analyzing public opinion of SNAP on social media and news outlets, as well as tracking Georgia politicians' voting records on issues relating to food insecurity.

This project focuses on utilizing natural learning processing tools, sentiment analysis, machine learning, and text mining to capture public opinion on the Supplemental Nutrition Assistance Program on a national and state level.

One objective of this project is to explore how discourse regarding SNAP varies geographically. While the ACFB has hypotheses based on their experiences, they do not have any quantitative measures to support their conjectures as of yet. After analyzing the sentiment of the data gathered from social media and news outlets, spatial analysis was used to identify geographic variation in SNAP sentiment.

In addition to better understanding public opinion on SNAP, the ACFB is also interested in the the voting records of Georgia politicians in Congress and in the Georgia General assembly. Having easy access to representatives' voting records on bills regarding food insecurity will help the food bank prepare for policy meetings with these politicians. Ultimately, this research will produce a tool that communicates dominant narratives and opinions about SNAP so that the ACFB Advocacy team can better communicate to stakeholders about SNAP. This research was conducted in conjunction with the Atlanta Community Food Bank and the Data Science for Social Good program at the Georgia Institute of Technology.

## 2.METHODS
### 2.1.DATA COLLECTION
Tweets were collected for a one-month period using the streamR package in R, which accesses the Twitter Streaming API. The Streaming API allows access to around one percent of tweets that are being tweeted in real time (Morstatter, Pfeffer, Liu, & Carley, 2013).

The collection of the tweets was based on search terms related to SNAP: "SNAP," "food stamp," "food stamps," and "EBT." The tweets were selected if they had any meaningful content regarding SNAP and were further sorted based on if they were geotagged. There were approximately 700 tweets about food stamps that were used for this analysis.

Finally, the voting records of Georgia state representatives were collected through Open States, a site that collects data on state representatives. Bills were selected if they contained the phrases "food stamps", "SNAP", "food bank", "food desert," "hunger," "food insecurity," or "georgia peach card". Bills with no votes were removed, and votes by representatives no longer in office were removed.

## 2.2.TEXT MINING AND SENTIMENT ANALYSIS

Sentiment analysis was used to assess the discourse regarding SNAP. Sentiment analysis is a form of text analysis that determines the subjectivity, polarity (positive or negative) and polarity strength (weakly positive, mildly positive, strongly positive, etc.) of a text (Liu, 2010). In other words, sentiment analysis tries to gauge the tone of the writer.

There are two main approaches in classifying the sentiment of a given text: supervised classification and unsupervised classification. Supervised classification requires labeled data and its features must be extracted from the data. Examples of features are part of speech tags, most frequent words, reading level, and name entity tags. Labels are nominal data. With these features and labeled data, any type of supervised learning approach can be used. It creates a model that is suitable for the data set with the label, so that it can predict with a new dataset without the label. This model is totally dependent on the dataset and its characteristics. When the characteristics in the dataset are similar, supervised learning classification tends to perform well. This applies for the Twitter data set, where the length and diction of the tweets are similar to one another. For the Twitter data, the scikit-learn package from Python was used to perform supervised classification (McKinney, 2010; Pedregosa et al., 2011).

Unsupervised classification was performed on the news articles. Unsupervised classification is different from supervised learning where the model is independent from the data, but it follows specific rules that it has in place. In this case, it uses a pre-existing lexicon, a dictionary that contains more information than just its meaning, and syntactic data, set of rules regarding the syntax of the sentence structure, to determine its sentiment. This method creates a numerical value or a probability of the sentiment rather than a nominal classification. This form of classification was used to analyze the news articles because the text has varying length, style, dictions, and form depending on the writer, which requires a bigger dataset to perform supervised classification.

The Vader and AFINN packages in Python were used to conduct unsupervised sentiment analysis. Vader is short for Valence Aware Dictionary Sentiment Reasoner,and is a lexicon and rule-based sentiment analysis tool (Hutto & Gilbert, 2014). AFINN is a dictionary of words that rates connotation severity from -5 to 5 (Årup Neilsen, 2011). The actual sentiment score was given as the sum of the word score within a sentence. The Vader tool gauges the overall syntactical sentiment more so than the word usage. Conversely, AFINN gauges the type of words that are being used and their intensity. Additionally, sentences with key words (words relating to SNAP) were given a higher weight so that sentiment towards this issue would be amplified.

Each article was tokenized to the sentence level, and each sentence was given a sentiment score according to the two sentiment analysis tools (NLTK documentation, 2016). Then, the scores were aggregated for each article with the weight that was assigned to each sentence. This aggregated score represents the sentiment of the article. To take into account of impact of the article, each article was then aggregated in regard to the traffic level of the website and the reading level of the article (Bansal, 2015). This process is visualized in Figure 1.

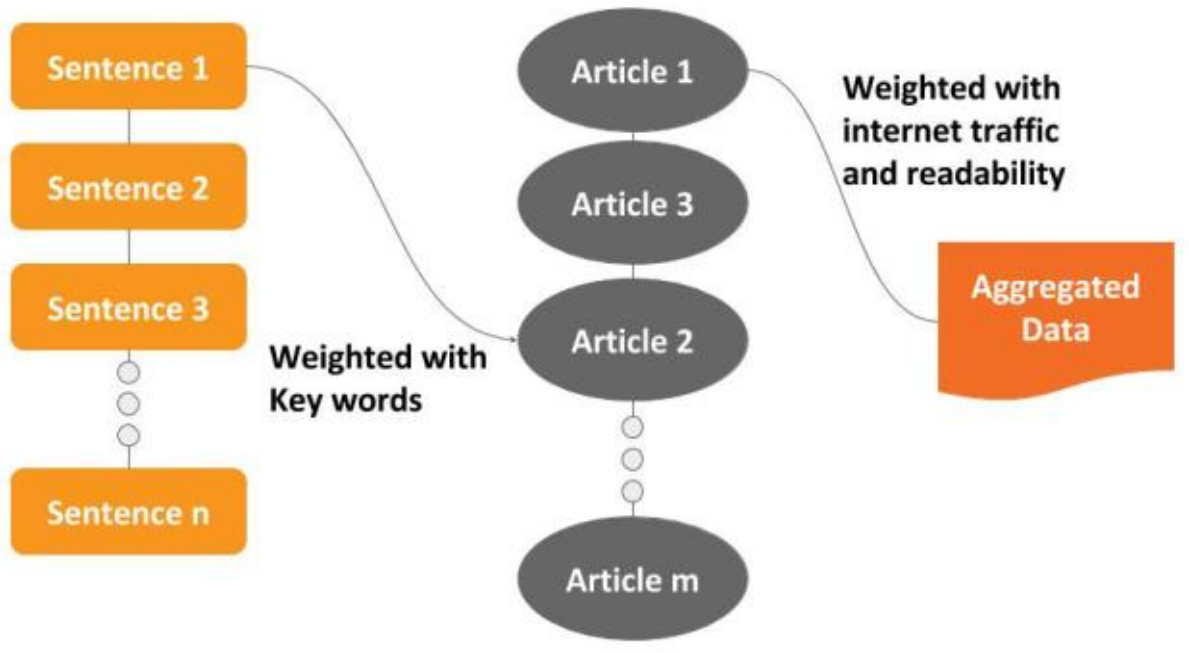

**Figure 1: Sentiment Analysis Methods**

Additionally, information on the arguments and topics in these articles would be very useful to the ACFB. To do this, preliminary topic modeling (Latent Dirichlet Allocation) has been performed to extract the topical words from the set of text. It returns a set of words with probabilistic weight on each of the word to indicate its importance. Bigram collocation has been used to detect sets of two words that are most frequent and meaningful. Term frequency inverse document frequency (TFIDF) was used to detect important words across all the documents. Name Entity Recognition (NER) from the Stanford Natural Language Processing Group (Finkel, Grenager, & Manning, 2005) and genism (Řehůřek & Sojka, 2010) were used to detected key people or locations mentioned in the articles. After generating all the statistics, each word within TFIDF, bigram collocation and NER was multiplied with the weight that was computed with each of the documents.

Then, all the words were aggregated into a list. Using this list, a word cloud can be generated to visualize meaningful words. Word clouds are especially of interest to our partners at the food bank. Along with the word cloud, its aggregation by each date will help the viewer understand the subject of the sentiment to better decipher the public opinion about SNAP.

## 2.3.SPATIAL ANALYSIS

The AFINN and Vader scores were linked to the geocoded outlets. Using ArcMap 10.4, spatial analysis was conducted on the outlets to determine whether there was any clustering of articles that had positive or negative sentiment about SNAP. In order to do this, a hexagon grid was created over the extent of a U.S. shapefile and a spatial join was conducted in order to join the number of news outlets to the hexagon polygons. After the spatial join, hot spot analysis was done by calculating the Getis-Ord Gi* statistic. The Getis-Ord Gi* statistic determines where there is clustering of cold spots and hot spots though looking at the location of features in relation to neighboring features (ESRI, 2017). Significance is determined based on looking at the proportion of the local sum of features and its neighbors to all the features (ESRI, 2017). If the difference between the calculated sum and the expected sum is very large, then the z-score is statistically significant (ESRI, 2017). In the context of this research, hot spots are areas in which the articles have a positive sentiment on SNAP and cold spots are areas in which the articles have a negative sentiment on SNAP.

# 3.RESULTS

## 3.1.SENTIMENT ANALYSIS

The results of the two sentiment analysis tools, Vader and AFINN, had very high correlation as classification, but the magnitude varied. There are many cases where Vader would score an article as having a positive sentiment while AFINN would score a negative sentiment. This may have been due to the fact that AFINN does not correctly account for syntactic information that may negate the meanings of words (such as "no" or "not). In the future, altering and refining these analysis tools to take this into account could generate more accurate results.

Although there was not a clear association between the features extracted from the text and its sentiment score, a strong correlation existed between extreme sentiment and extreme media bias. Articles with extreme right bias tended to have extreme sentiment scores while articles with extreme left bias tended to have relatively less extreme sentiment scores. Higher traffic news websites' sentiment correlated with the current events about SNAP (Figure 2). In May of 2017, Trump's budget was released; this budget included a large proposed cut on SNAP. In regard to this event, articles had negative sentiment scores. This trend negative continued as editorials on the budget cut were written.

One important way to interpret these sentiment scores is that they are gauging people's sentiment when the text was written. When gauging opinion on certain topic, the sentiment analysis can be very misleading. It is extremely rare for a speaker to comment directly about the food stamps program itself. For example, the negative sentiment during the Trump food stamp cuts were mostly written by people supportive of the program itself. To effectively use the sentiment analysis tool, one must look at the titles and key words of the articles that were scored a certain way. The visualization on the application will provide a quick overview of sentiments overtime along with major events.

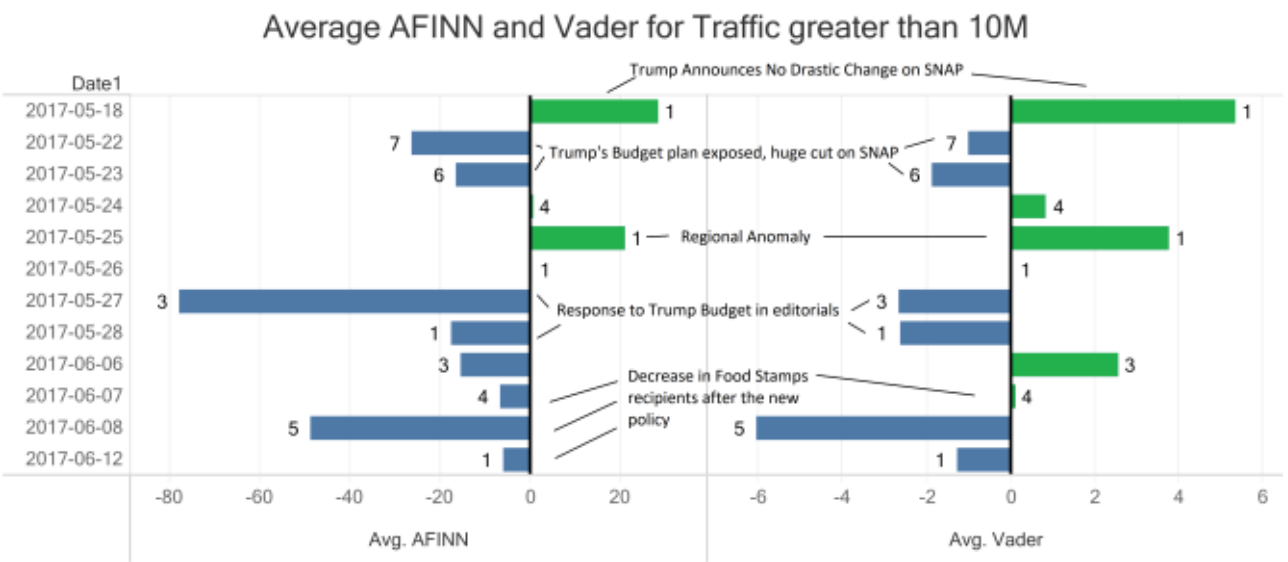

**Figure 2: This figure shows the difference in sentiment scores over the month of May and June 2017, as analyzed by AFINN and Vader. Green indicates a positive score and blue indicates a negative score. Numbers beside the bars indicate count of the articles. Each bar and its sentiment is matched with the corresponding current event.**

## 3.2.SPATIAL ANALYSIS

Based on the hot spot analysis that was conducted on the AFINN sentiment scores of 1,250 of the 2,239 news outlets, the news outlets with negative AFINN scores were more concentrated compared to the news outlets that had positive AFINN scores. Many of the news outlets that have a negative sentiment on SNAP were in the Midwest, especially in Indiana, Michigan, and Illinois. On the other hand, news outlets with positive AFINN scores were more dispersed, with a concentration of positive AFINN scores in the South and Southeast. This could be due to the high enrollment of individuals on SNAP such as in the District of Columbia, Mississippi, and Tennessee which have the highest number of individuals on SNAP in the nation (Rawes, 2015).

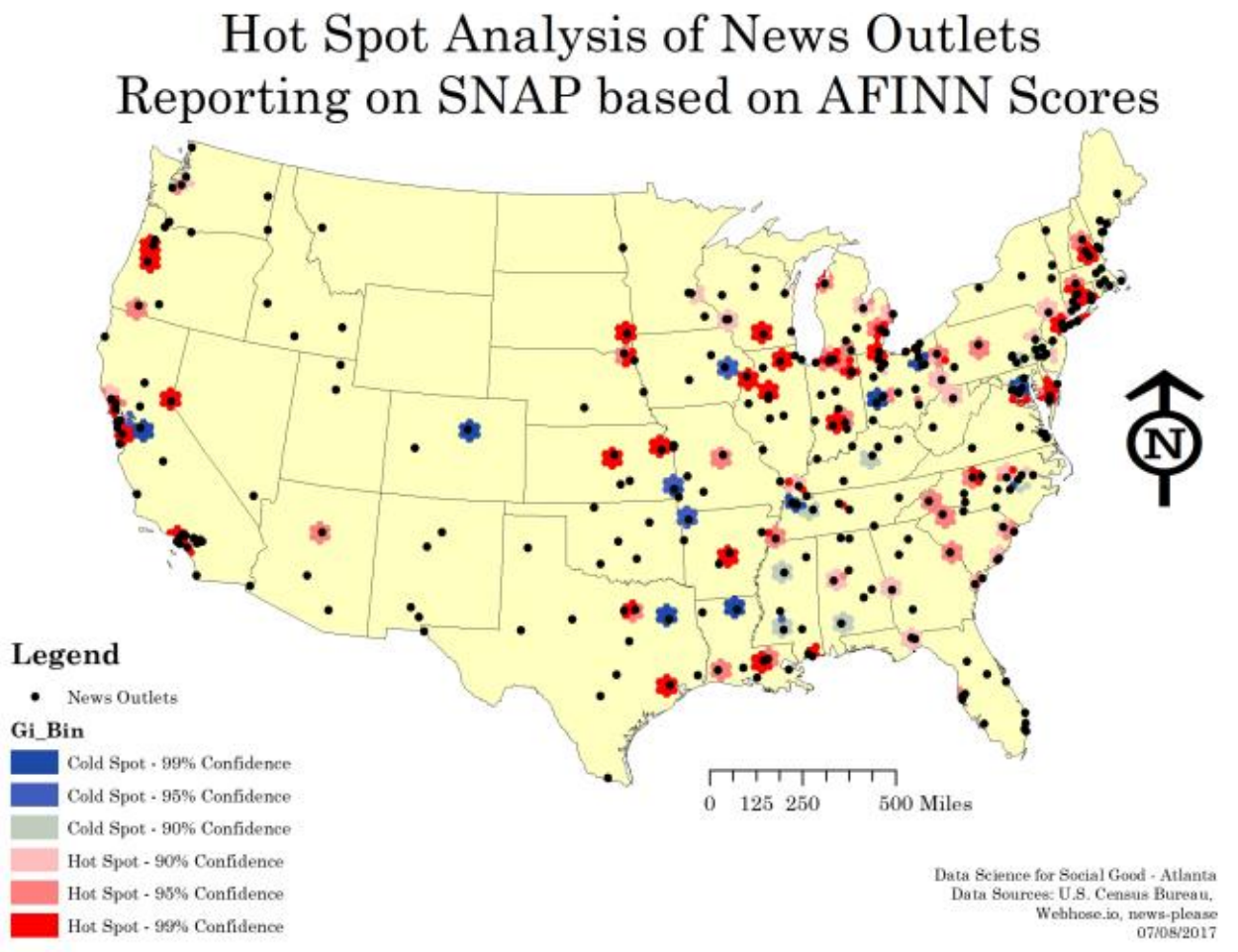

**Figure 3: Hot spot analysis of the AFINN sentiment scores of news outlets reporting on SNAP. Cold spots indicate a concentration of outlets with negative AFINN scores and hot spots indicate a concentration of outlets with positive AFINN scores.**

# 4.DELIVERABLES

**4.1.**The results of the sentiment analysis, text mining, and aggregation of voting records will be contained in an on-line application which was created using the Shiny web framework in R. This application will allow the ACFB to better understand reporting and public opinion on SNAP through interactive visualizations such as world clouds, maps, charts, and graphs. The "Background" section of the application gives an overview of the SNAP program and the importance of the program in various contexts. "The Word on SNAP" section will provide visualizations of how SNAP is discussed in social media and media outlets, such as the interactive word cloud that is displayed in Figure 4.

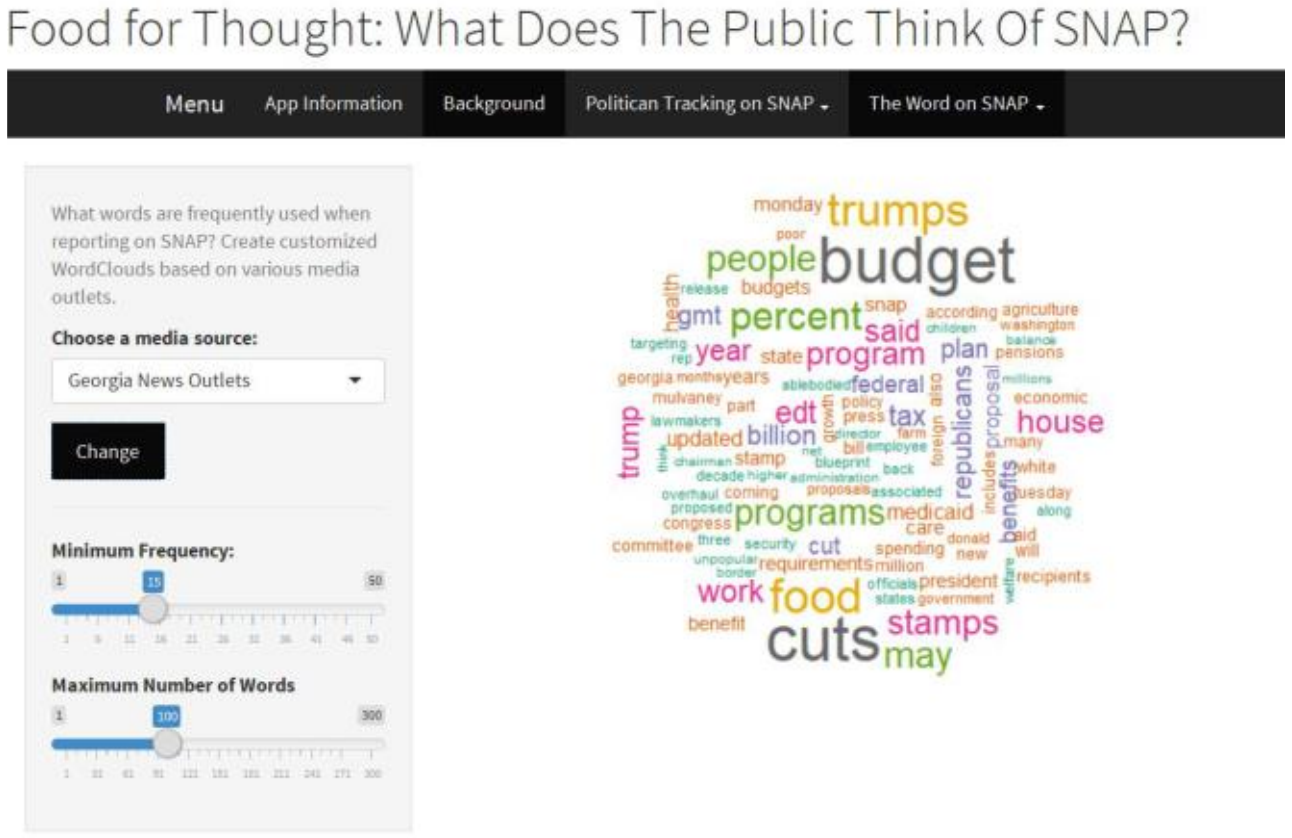

**Figure 4: An interactive word cloud showing the most frequent words found in twenty conservative news outlets.**

This section also includes an interactive map called "SNAP InfoMap" in which users can see the location and types of news outlets reporting on SNAP and the affiliated sentiment score attached to each outlet.Users are also able to explore how the location of the news outlets correlates to the socioeconomic characteristics that are related to the program such as the percentage of households that are on SNAP.

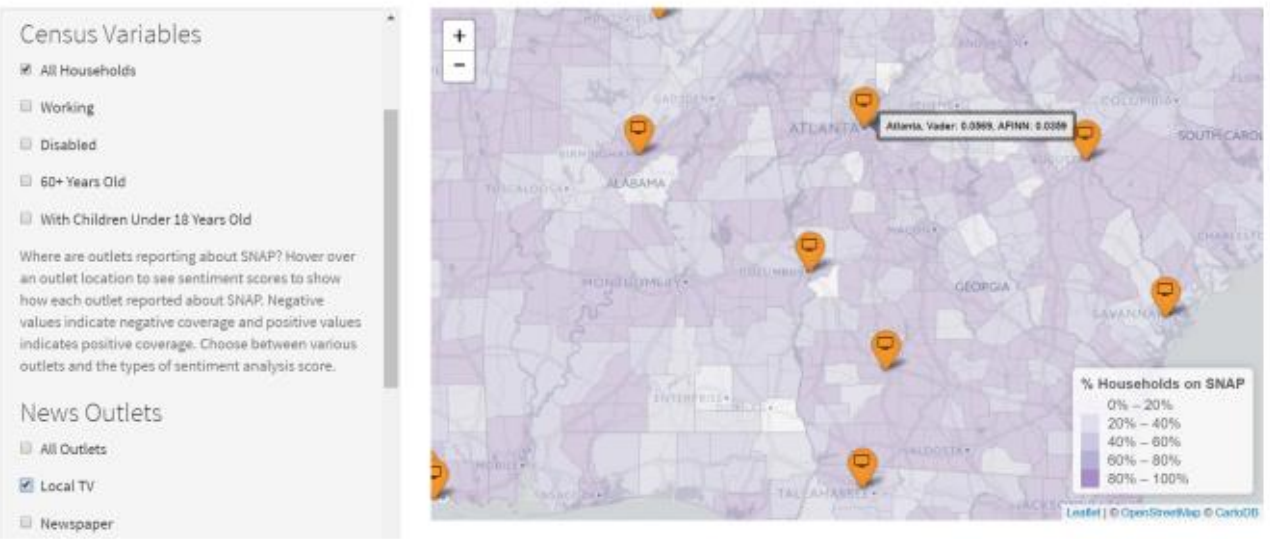

**Figure 5: The SNAPInfo Map is an interactive map interface which allows users to explore how positive and negative coverage on SNAP varies by media outlet, location, and socioeconomic factors.**

In addition to the word cloud and the interactive map, a sentiment analysis tool was created to show the average AFINN and Vader scores for the news outlets and tweets and how the sentiment on SNAP changes through a specified time period. For example, when President Trump announced a budget cut on SNAP, most of the articles for higher trafficked websites had a negative sentiment score, as indicated in Figure 6.

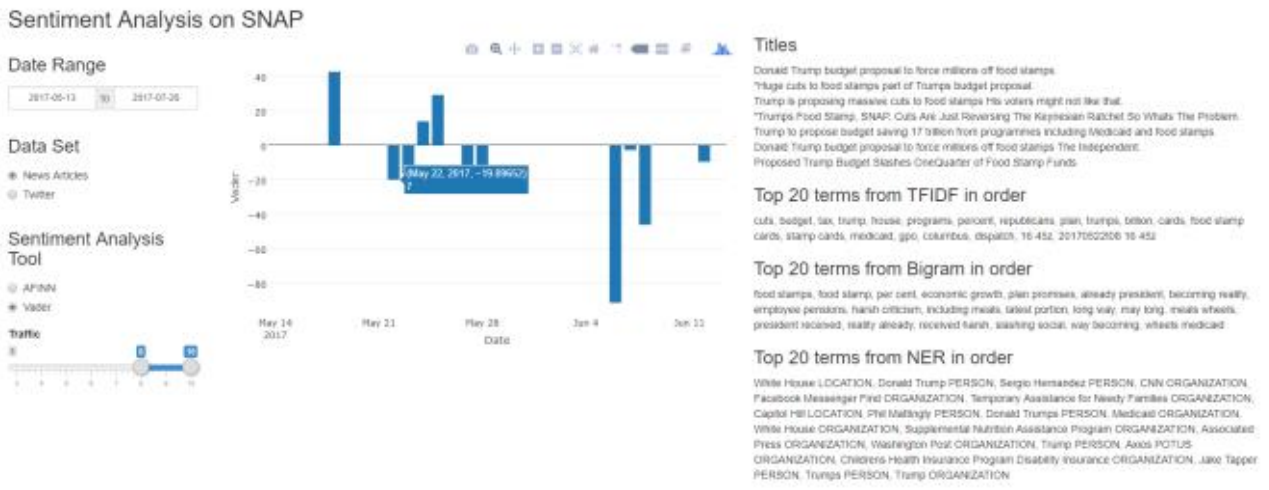

**Figure 6: Sentiment Analysis Tool**

The "Politician Tracking on SNAP" section will allow one to look up the voting record of Georgia legislators on bills related to SNAP on the state level, as shown in Figure 7. The word cloud uses TFIDF in order to show which words are prominent in a set of articles which is related to the size of the word in the visualization.

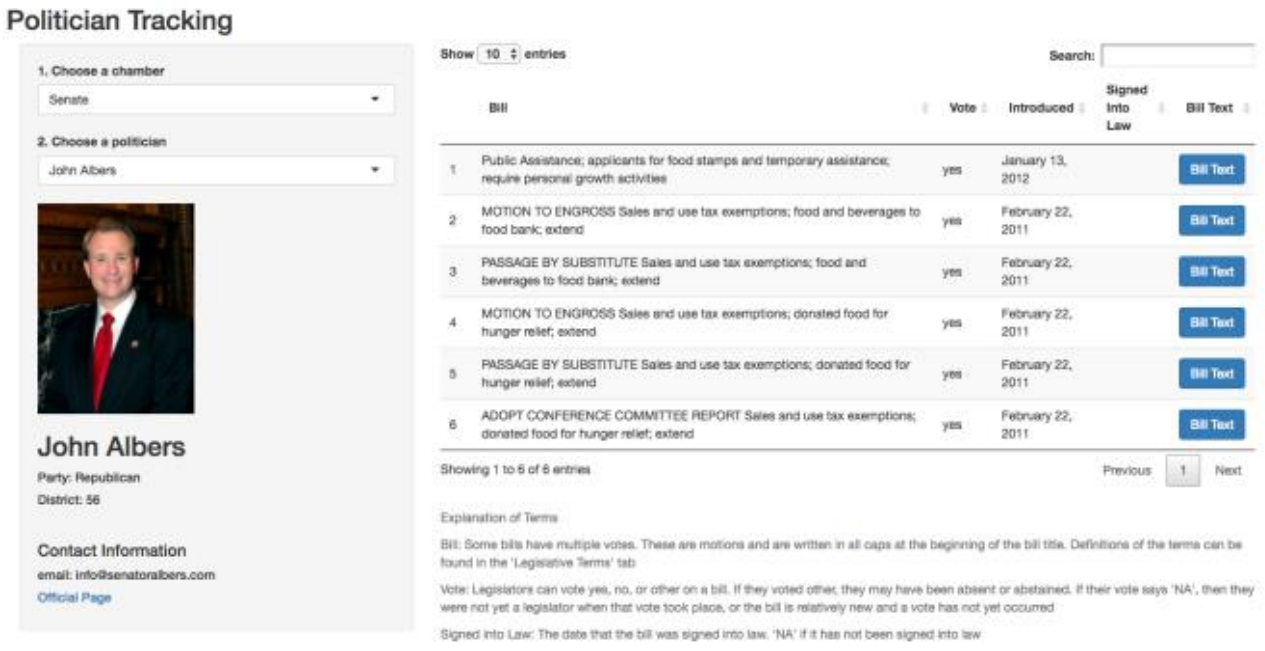

**Figure 7: Politician Tracking Tool**

# 5.LIMITATIONS

While this research shows the potential of using data science techniques to explore the various discourses regarding SNAP, there are limitations in using these techniques. For example, a sentiment score that indicates that an article is negative does not tell us if the article is negative in regards to SNAP. Furthermore, it does not indicate that the article contains information analyzing critical threats made to the SNAP program. It is impossible to exactly fact check the sentiment towards SNAP with this tool. However, connecting the sentiment with actual words associated from the article will help discern the meaning of the sentiment score. Additionally, different aggregation techniques and different datasets can yield to different results. Exploring different types of aggregation, such as grouping progressive article and conservative articles separately to refine the sentiment scores would be beneficial.

The news articles that were collected for this study were limited to articles that were published within the last 30 days from the time of collection. In order to perform a better sentiment analysis, more historical articles from the web must be scraped to see the trend of the sentiment. This will be useful in comparing the values computed during past events that affected SNAP. Scraping from webhose.io moving forward will create a richer dataset to work with. Similarly, the Twitter dataset was very limited to perform machine learning on. if we were to collect more tweets as time passes, it may be valuable to see how sentiment towards SNAP is moving and how it actually compares to the two sentiment analysis tools that were used. Another limitation of the machine learning is the criterion on labeling the sentiment as well as inevitable bias in labeling the data. These limitations should be discussed as the project moves forward.

Another limitation has to do with the tools that were used to perform sentiment analysis. Vader was originally created for Twitter data, which has different text features than news articles. Although the articles were tokenized to the sentence level to increase the precision, the Vader model could have produced less accurate results. Similarly, the lexicon that is being used for AFINN is limited to 2477 words. A larger lexicon would allow for more accurate results. Additionally, AFINN doesn't consider the syntax of the sentence. Adding a negation of the word depending on the key words could drastically improve the result. Trying supervised classification with the pre-existing corpus and labeled data could result in different findings.

# 6.CONCLUSION

This research took a novel approach to gauging public opinion on SNAP. Commonly, public sentiment is gauged through a poll or survey as opposed to using more exploratory methods such as sentiment analysis, text mining, and spatial analysis. While this study was heavily based on data science techniques, what truly drove the direction of the study was the collaboration between the ACFB. Through regular meetings with the ACFB, we were able to get feedback on whether the project was going in the direction that they wanted and made changes accordingly in terms of technique and creation of visualizations.

The food bank is using our tools to inform their interaction with media outlets, to prepare for meetings with politicians, and to adjust their social media and outreach messaging. For example, the Director of Government Affairs used the app in preparation for a meeting with a congressman. When the congressman talked about SNAP during the meeting, the Director of Government Affairs tracked his word usage to see how it compared to positive and negative arguments presented in the tools. Through an iterative process, we were able to apply data science techniques to help the ACFB fulfill their organizational goals.

## 7.ACKNOWLEDGEMENTS

We thank our mentor, Carl DiSalvo, Associate Professor and Coordinator for the MS in Human-Computer Interaction at Georgia Tech for his guidance and advice.

We also thank our Food Bank partners Lauren Waits, Director of Government Affairs; Allison Young, Marketing Manager; and Jocelyn Leitch, Data and Insights Analyst; for educating us about food policy, food insecurity in Atlanta and across the nation, and the work of the Atlanta Community Food Bank.

Finally, we would like to thank the staff and students participating in the Data Science for Social Good - Atlanta program for their support and assistance.

## 8.REFERENCES